\documentclass[sn-mathphys]{sn-jnl}

\theoremstyle{plain}
\newtheorem{thm}{Theorem}[section]

\theoremstyle{definition}

\newtheorem*{thm*}{Theorem}

\theoremstyle{remark}

\jyear{2021}%

\theoremstyle{thmstyleone}%
\theoremstyle{thmstyletwo}%

\theoremstyle{thmstylethree}%

\begin{document}

\title[MOL]{Method of lines for valuation and sensitivities of Bermudan options}


\author[1]{\fnm{Purba} \sur{Banerjee}}\email{purbab@iisc.ac.in}

\author[2]{\fnm{Vasudeva} \sur{Murthy}}\email{vasu@tifrbng.res.in}
\equalcont{These authors contributed equally to this work.}

\author*[3]{\fnm{Shashi} \sur{Jain}}\email{shashijain@iisc.ac.in}
\equalcont{These authors contributed equally to this work.}

\affil[1]{\orgdiv{Department of Mathematics}, \orgname{Indian Institute of Science}, \orgaddress{\city{Bangalore}, \postcode{560012},  \country{India}}}

\affil[2]{\orgname{TIFR Centre For Applicable Mathematics}, \orgaddress{\city{Bangalore}, \postcode{560065},  \country{India}}}

\affil*[3]{\orgdiv{Department of Management Studies}, \orgname{Indian Institute of Science}, \orgaddress{\city{Bangalore}, \postcode{560012},  \country{India}}}


\abstract{In this paper, we present a computationally efficient technique based on the \emph{Method of Lines} (MOL) for the approximation of the Bermudan option values via the associated partial differential equations (PDEs). The MOL converts the Black Scholes PDE to a system of ordinary differential equations (ODEs). The solution of the system of ODEs so obtained only requires spatial discretization and avoids discretization in time. Additionally, the exact solution of the ODEs can be obtained efficiently using the exponential matrix operation, making the method computationally attractive and straightforward to implement. An essential advantage of the proposed approach is that the associated Greeks can be computed with minimal additional computations.  We illustrate, through numerical experiments, the efficacy of the proposed method in pricing and computation of the sensitivities for a European call, cash-or-nothing,  powered option, and Bermudan put option.  }




\maketitle

\section{ Introduction}

Pricing an option with early exercise features, such as Bermudan options, is a problem of practical importance with numerous pricing methods proposed in the literature, each with its own set of advantages. When fast and accurate evaluation of the option price is the objective, the Fourier based methods are commonly used numerical techniques. Some of the proposed Fourier based pricing methods include the COS method \cite{fang2009novel}, data-driven COS \cite{ddCOS}, and fast Fourier transform-based approach \cite{carrMadan}. Monte Carlo based schemes are popular for pricing the early exercise options under multidimensional stochastic processes.  The traditional lattice-based methods are often impractical in such cases due to the curse of dimensionality. Popular simulation-based pricing models include Longstaff and Schwartz method \cite{LSM}, stochastic mesh method by Broadie and Glassermen \cite{SMM}, and stochastic grid bundling method \cite{SGBM}. There has been an increasing interest in neural network-based pricing methods for early exercise options. Some of the recent ones include \cite{oosterleeDeep}, \cite{RLNN}, and \cite{becker2019deep}.

For low dimensional options, often finite difference methods are used for approximating the solution to the underlying partial differential equation, as approximate option prices for a grid of underlying values (see for instance \cite{kees}) can then be obtained. Solving efficiently the partial differential equations (PDEs) for American styled options using finite difference schemes have been extensively studied in \cite{brennan}, \cite{adi}, and \cite{adi1}. The ability to compute prices along a spatial and temporal grid of underlying stochastic states is helpful for the simulation of future exposure. Future exposure is required, for instance, for credit valuation adjustment or for the risk management purposes, such as determining the potential future exposure of a particular position. 

In this paper we propose an accurate and efficient approach based on the Method of Lines (MOL) to compute the value of Bermudan and European options for a grid of underlying values at any time prior to maturity. The MOL for solving evolutionary PDEs consists of two parts, first involves discretization of the space variable and then writing the PDE as a system of ordinary differential equations (ODEs). The second part involves efficiently solving the system of ODEs so obtained. The main advantage of the proposed approach is that it solves the ODEs so obtained exactly, by writing the solution to the ODE as an exponential of a matrix. As most modern numerical packages come  with  efficient solvers for computing the exponential of matrices, we can efficiently compute the exact solution for the ODE. The above is possible in the Black-Scholes framework as the coefficients of the PDE are time independent and the PDE is linear.

 Several researchers have worked on the MOL in the past few decades, most notable among which include  \cite{hamdi2007method},  \cite{schiesser2009compendium}, and \cite{lee2003ordinary} who provide a general elucidation to the method. For a detailed discussion on the application of time discrete MOL for pricing of options the readers can refer \cite{meyer2014time}.  Similar time-discretised approaches have been implemented for pricing of American put option in the Black-Scholes framework \cite{mol2}, for put options under jump-diffusion dynamics \cite{meyer1998numerical}, call options under stochastic volatility \cite{adolfsson2013representation}, American options with stochastic volatility and  interest rates \cite{kang2014pricing}, American options under a regime-switching GBM  \cite{chiarella2016pricing}. The method proposed here is closely related to the approach proposed in \cite{mol1}, and \cite{mol2}.
 
 Some of the recent regulatory requirements, for instance exchange of initial margin computed based on the ISDA standard initial margin model (SIMM), requires efficient computation of the sensitivities of the derivative with respect to its underlying risk factors. Further, in order to compute the associated margin valuation adjustment (MVA), to manage the funding costs for posting initial margin over the lifetime of a derivative, sensitivities along the paths are required (see \cite{jain2019rolling}). Computation of the sensitivities along the paths, for all the underlying risk factors, poses noteworthy computational challenges. 
 
 Some of the key features of the study presented in this paper are:
 \begin{itemize}
 \item We provide the exact solution for the ODEs obtained using MOL from the Black Scholes PDEs for European options.
 \item The scheme avoids discretization in the time domain and the solution is closed form in time. Therefore, one can obtain the approximate option price at any time instant prior to the maturity of the option.
 \item The matrix expressions for the Greeks: delta, theta, gamma, and vega are obtained which allows evaluating the various sensitivities at the spatial grid points with a few additional computations. This feature makes the approach attractive when sensitivities along scenarios are required. 
 \item  We extend the approach to Bermudan options, by formulating a Bermudan option as a sequence of European options with appropriate pay-offs.
 \end{itemize}

  The paper is organized as follows:
   We first begin with the formulation of the problem and defining the notations used in Section \ref{pf}.  Section \ref{md} describes in detail the methodology used. In Section \ref{gk}, the expressions for the  Greeks are obtained. Section \ref{nl} presents detailed numerical examples for European call, cash or nothing, powered option, and Bermudan options to illustrate the efficiency of the proposed method. Finally, we provide some conclusions  in Section \ref{cl}.
   
\section{Problem formulation}
\label{pf}
A Bermudan option is defined as an option where the buyer has the right to exercise at a set of discrete time points. We denote the exercise times by $$\mathcal{T} \equiv \{t_1,\ldots,t_e,\ldots,t_E=T\},$$ where, $0 = t_0 \leq t_1 \leq t_2 \ldots t_{E-1} \leq T,$  $T$ is the maturity date of the option, and $E$ denotes the number of early exercise opportunities.  For ease of notation, we assume that the exercise dates are equally spaced, i.e., $t_{e+1} -t_e = \Delta t$.
 
The pay-off received by the holder of the option upon exercising his rights at the opportunity $t_e,\ e \in \{1,\ldots, E\}$  is given by $\phi(S_{t_e}).$ The continuation value, or the value of the option if the holder decides to not exercise and continue holding the option at $t_e,$ is defined as:

$$
c(t_e,S_{t_e}) = e^{-r\Delta t} \mathbb{E}\left[U\left(t_{e+1},S_{t_{e+1}}\right)\mid \mathcal{F}_{t_e}\right],
$$

where $r$ is the risk free rate (we assume it to be constant), $\mathcal{F}_{t_e}$ is the filtration associated with the stochastic process $S_{t},$ the expectation is taken under the risk-neutral measure, and $U\left(t_{e+1},S_{t_{e+1}}\right),$ is the option value function at $t_{e+1}.$

The option value at any time $t$ is then solved using the following dynamic programming formulation. 
The value of the option at the terminal time T is given by,
$$
U\left(t_{E},S_{t_{E}}\right) = \max\left(\phi(S_{t_E}),0\right).
$$

Recursively, moving backwards in time, the following iteration is then solved. Given $U\left(t_{e+1},S_{t_{e+1}}\right),$ has already been resolved,
\begin{eqnarray}\nonumber
c(t_e,S_{t_e}) &=& e^{-r\Delta t} \mathbb{E}\left[U\left(t_{e+1},S_{t_{e+1}}\right) \mid \mathcal{F}_{t_e}\right]\\
U\left(t_{e},S_{t_{e}}\right) &=& \max\left(c(t_e,S_{t_e}), \phi(S_{t_e})\right).
\end{eqnarray}
The value of the option at $t_0,$ assuming there is no exercise opportunity at $t_0$ is then given by  $U(t_0, S_{t_0}) = c(t_0, S_{t_0})$

\section{Methodology}
\label{md}
   Under the Black-Scholes framework \cite{black1973valuation}, when the  underlying stock follows the geometric Brownian motion (GBM), the price of the European option, $U,$ satisfies the following PDE,
   \begin{equation}\label{bs}
      \frac{\partial U}{\partial t}+\frac{1}{2}\sigma^2 S^2 \frac{\partial^2 U}{\partial S^2}+rS\frac{\partial U}{\partial S} -rU=0,
    \end{equation}
 where the underlying $S$ follows the GBM process, 
   $$
   dS_t = \mu S_t + \sigma S_t dW_t,
   $$
   
  $r$ is the risk free rate, $\mu$ is the drift of the underlying process, $\sigma$ is the implied volatility,  and $W_t$ is the standard Brownian motion.

     In order to numerically solve (\ref{bs}) for $t \in [0, T],$ we first transform it into the familiar forward in time parabolic partial differential equation by replacing $U$ with $u$, $S$ with $x$ and $t$ with $\tau= T-t$, thereby obtaining the following initial-boundary value problem for $u(\tau, x),$
     
   \begin{align}
   \label{bs2}
       \frac{\partial u}{\partial \tau}=\frac{1}{2}\sigma^2 x^2 \frac{\partial^2 u}{\partial x^2}+rx\frac{\partial u}{\partial x} -ru, ~~x>0,~~\tau \in (0,T],
   \end{align}
   
   with the initial condition given by Equation \ref{icbEU}.
      
   \begin{equation}\label{icbEU}
   u(0,x)\equiv u_0(x) = \phi(x).
   \end{equation}
    
 The value of Bermudan option, with early exercise dates $\mathcal{T} \equiv \{t_1,\ldots,t_e,\ldots,t_E=T\},$ can be formulated as follows. Define $\tau_e = T-t_{E-e},$ for $e = 0,1,\ldots,E,$ such that $\tau_0 = 0$ and $\tau_E = T.$ The fair value function $u$ of Bermudan option satisfies the PDE \ref{bs2} with natural boundary condition on each time interval $\left[\tau_{e-1}, \tau_{e}\right),$
for $e=1,2,\ldots,E.$ The initial condition for $\tau_0$ is 

$$
u(\tau_0,x) = \phi(x),
$$

while the initial condition for solving the PDE \ref{bs2} in the interval  $(\tau_{e-1}, \tau_e), \, e=2,\ldots,E$ is given by

\begin{equation}\label{icb}
u(\tau_{e-1}, x) = \max\left(\phi(x), \lim_{t\uparrow \tau_{e-1}} u(t, x)\right)
\end{equation}   

 Condition (\ref{icb}) is non-linear and arises from the early exercise feature of Bermudan options and represents the optimal exercise condition.  We solve the PDE \ref{bs2} in the interval $[\tau_{e-1}, \tau_{e}),$  to obtain the initial condition for the subsequent interval, i.e. $[\tau_{e},\tau_{e+1}).$

   Since numerically we cannot work with the infinite spatial domain $[0,\infty)$, the spatial domain is truncated at $x=L.$ We use a non-uniform grid in the spatial dimension, where a family of uniform grid $\eta_n = \frac{n}{N}, \, n \leq N,$ ($N$ is the number of discretization points)  defined in the interval $[0,1]$ generates a two parameter family of quasi-uniform grid 
   $$x_n = c\frac{\eta_n}{d-\eta_n},$$ 
   where $\eta_n \in [0,1], x_n \in [0, \infty).$ This map has two control parameters $c>0,$ and $ d > 1$ (see  \cite{fazio2014finite}).  Once a uniform mesh has been generated in $[0,1]$, we shall use the above mentioned transformation to generate a non-uniform grid with $x_0 =0,$ and truncate $x_{N+1}$ at $L$. Following this, we set $x_i -x_{i-1} =h_i$, for $i=1....N$ and $h_{N+1}= L- x_N.$ We then reduce (\ref{bs2}) to a set of ODEs by approximating the spatial derivatives in the following manner (see \cite{volders2014stability}):  at each grid point $x_i$, we approximate the option value by
   \begin{align*}
       u(\tau, x_i)\approx U_i (\tau).
   \end{align*}
   Further, the approximation of the first derivative is given by,
   \begin{align}
   \label{ux}
       u_x(\tau, x_i)\approx\frac{U_{i+1}(\tau) -U_{i-1}(\tau)}{h_i +h_{i+1}}= D U_i(\tau).
   \end{align}
   Analogously, the approximation for the second derivative is
   \begin{align}
   \label{uxx}
       u_{xx}(\tau, x_i)\approx \frac{2 U_{i-1}(\tau)}{h_i (h_i +h_{i+1})}-\frac{2 U_i(\tau)}{h_i h_{i+1}}+\frac{2 U_{i+1}(\tau)}{h_{i+1} (h_i +h_{i+1})}=D^2 U_i(\tau).
   \end{align}
   It should be noted that for a uniform mesh we have $h_i =h$, hence the above approximations would in the case of uniform mesh coincide with the familiar second order approximations to the first and second derivatives, i.e.,
   \begin{align*}
       u_x(\tau, x_i)\approx\frac{U_{i+1}(\tau)-U_{i-1}(\tau)}{2h}, ~~u_{xx}(\tau, x_i)\approx \frac{U_{i+1}(\tau)-2U_i(\tau) +U_{i-1}(\tau)}{h^2}.
   \end{align*}
   Upon replacing the spatial derivatives in (\ref{bs2}) by the approximations stated above, we obtain the following system of ODEs,
   \begin{align}
   \label{eq}
       \frac{d U_i(\tau)}{d \tau}=\frac{1}{2}\sigma^2 {x_i}^2 D^2 U_i(\tau) +r x_i D U_i(\tau) -r U_i(\tau),~~\text{for}~~i=1,...N.
   \end{align}
   In order to obtain the value of the option, the matrices corresponding to the coefficients of the derivative approximations need to be defined. Consequently, the matrix corresponding to the second order derivatives is defined as follows,
   \begin{align*}
   A=
   \begin{bmatrix}
   \gamma_1 & \alpha_1 &...& 0\\
   \alpha_{-2} &  \gamma_2 & ...& 0\\
   ............\\
   0 &...........& \alpha_{-N} & \gamma_N
   \end{bmatrix}
    \end{align*}
   where,
   \begin{align*}
       \gamma_j =-\frac{\sigma^2 {x_j}^2} {h_j h_{j+1}},~~j=1,...N,\\
       \alpha_j =\frac{\sigma^2 {x_j}^2 }{(h_j+ h_{j+1})h_{j+1}},~~j=1,...N\\
       \alpha_{-j} =\frac{\sigma^2 {x_j}^2} {(h_j+ h_{j+1})h_{j}}~~j=1,...N
   \end{align*}
   and the matrix corresponding to the first order derivatives is
   \begin{align*}
        B=
   \begin{bmatrix}
   0 & b_1 & ....& 0\\
   -b_2 & ........\\
   .......&....& ...& b_{N-1}\\
   0 & .... &-b_N & 0
   \end{bmatrix}
   \end{align*}
   where,
   \begin{align*}
       b_j =\frac{r x_j }{h_j + h_{j+1}}, j=1,...N.
   \end{align*}.
   
   Finally the zero-th order term in (\ref{bs2}) is incorporated in the following matrix,
   \begin{align*}
   C=
   \begin{bmatrix}
   -r & 0 & ...& 0\\
   0 & .... &...\\
   ..&....&...&..\\
   0...&..&0 & -r
   \end{bmatrix}
   \end{align*}
   
   with the boundary conditions being given by,
 \begin{align}
 \label{fcond}
   F(\tau)=
   \begin{bmatrix}
   (\alpha_{-1} -b_1)u(\tau,0)\\
   0\\
   .\\
   .\\
   .\\
   0\\
   (\alpha_N +b_N)u(\tau,L)
   \end{bmatrix}.
\end{align}
Using these, the ODE (\ref{eq}) can be compactly written as,
\begin{align}
\label{eq2}
    \frac{d U}{d\tau}=AU +BU+CU+F(\tau).
\end{align}
For the interval $[\tau_{e-1},\tau_{e}),$ the initial condition would be,
\begin{align*}
    U(\tau_{e-1}) ={[u(\tau_{e-1},x_1),u(\tau_{e-1},x_2),....u(\tau_{e-1},x_N)]^\top}.
\end{align*}

\subsection{Exact solution to the ODEs}
Denoting $\zeta =A+B+C$, the ODE (\ref{eq2}) can be readily solved to obtain the following solution,
\begin{align}
\label{eq3}
    U(\tau)=e^{\zeta(\tau-\tau_{e-1})} U(\tau_{e-1}) +\int_{\tau_{e-1}}^{\tau} e^{\zeta(\tau-s)}F(s) ds,
\end{align}
where $\tau \in [\tau_{e-1},\tau_{e}).$ The above is possible based on the observation that $\zeta$ is not a function of time. 
The computation of the expression above requires the explicit definition of $F(s)$ from Equation \ref{fcond}. While $u(\tau,0)$ is in general known to be equal to $u(\tau_{e-1},0),$  $u(\tau,L)$ is often not known. Hence, we draw inspiration from \cite{kangro2000far}, and for large enough $L,$ approximate its value as $u(\tau_{e-1},L)$, which readily yields,
\begin{align}
\label{fcond2}
   F(\tau)=F=
   \begin{bmatrix}
   (\alpha_{-1} -b_1)u(\tau_{e-1},0)\\
   0\\
   .\\
   .\\
   .\\
   0\\
   (\alpha_N +b_N)u(\tau_{e-1},L)
   \end{bmatrix}.
\end{align}
Given, $F$ is a constant, the second term in (\ref{eq3}) can then be integrated exactly to obtain the final expression,
\begin{align}
    \label{eq4}
    U(\tau)=e^{\zeta(\tau-\tau_{e-1}) }U(\tau_{e-1}) +\zeta^{-1}(e^{ \zeta(\tau-\tau_{e-1})}-1)F.
\end{align}

In order to compute Equation \ref{eq4} one needs to compute the exponential of a matrix, which has efficient implementation in most modern numerical libraries (see Appendix \ref{A} for details on matrix exponential). Having an exact solution for any $\tau$ in the interval $[\tau_{e-1}, \tau_e)$ for all the spatial grid points makes this approach attractive for exposure computation, where as proposed in \cite{kees} finite difference methods can be combined with Monte Carlo based methods. In the section to follow, we extend the above approach to obtain the various sensitivities, which again can be combined with Monte Carlo methods to obtain sensitivities along scenarios (see for instance \cite{kees2}). 

\section{Greeks using MOL}
\label{gk}

 The Greeks of an option represent the sensitivity of the price of the option with respect to changes in underlying parameters involved in the definition of the option. The primary Greeks of interest include: delta $\Delta =\frac{\partial U}{\partial S},$ vega $ \nu = \frac{\partial U}{\partial \sigma},$ theta $\Theta =\frac{\partial U}{\partial t},$ and rho $   \rho=\frac{\partial U}{\partial r}.$ Amongst second order Greeks, gamma, $\Gamma =\frac{\partial^2 U}{\partial S^2}.$ is often computed. Based on the methodology described above, we present simple extensions to compute the Greeks, not just at the initial point, but for any $t$ at the generated spatial grid. 

As a direct consequence of the approximation (\ref{ux}), one obtains the matrix corresponding to the first order and second order derivatives as
\begin{align}
       \Delta(\tau) = DU(\tau)
   \end{align}
   and the value of Gamma($\Gamma$) is given by the matrix product,
   \begin{align}
       \Gamma(\tau)= D^2U(\tau),
   \end{align}
   
   where the matrix D is defined as:
   \begin{align*}
        D=
   \begin{bmatrix}
   0 & d_1 & ....& 0\\
   -d_2 & ........\\
   .......&....& ...& d_{N-1}\\
   0 & .... &-d_N & 0
   \end{bmatrix}
   \end{align*}
   with,
   \begin{align*}
       d_j =\frac{1} {h_j + h_{j+1}}, j=1,...N.
   \end{align*}
   
   and $U(\tau)$ is obtained using Equation \ref{eq4}.

  In order to calculate Theta($\Theta$) one can differentiate $U(\tau)$ in (\ref{eq4}) to obtain the following formula for $\Theta$,
  \begin{align}
    \frac{\partial U(\tau)}{\partial t}=-\zeta e^{\tau\zeta} U_0 - e^{\tau\zeta} F
  \end{align}

  For the calculation of Vega($\nu$), a direct application of  calculus theory to  (\ref{eq4}) and the identity $\zeta.{\zeta}^{-1} = I_N$  yields the partial differential of $U(\tau)$  with respect to $\sigma,$
  \begin{align}
  \label{vega}
      \frac{\partial U(\tau)}{\partial \sigma}=\tau e^{\tau\zeta}A'U_0-\zeta^{-1}A'{\zeta}^{-1}(e^{\tau\zeta}-I)F+\tau{\zeta}^{-1}e^{\tau\zeta}A'F +{\zeta}^{-1}(e^{\tau\zeta}-I)F'
  \end{align}
   where, $A'$ and $F'$ are the matrices obtained by term-by-term differentiation of the matrices $A$ and $F$ (\ref{fcond2}) respectively with respect to $\sigma$.
   
   Analogously one obtains the formula for Rho($\rho$) ,
   \begin{align*}
       \frac{\partial U(\tau)}{\partial r}= \tau e^{\tau\zeta}(B' + C')U_0 -\zeta^{-1}(B'+C'){\zeta}^{-1}(e^{\tau\zeta}-I)F+\tau{\zeta}^{-1}e^{\tau\zeta}(B'+C')F
   \end{align*}
   \begin{align}
   \label{rho}
       + {\zeta}^{-1}(e^{\tau\zeta}-I)F_r'
   \end{align}
   
   where $B'$, $C'$ and $F_r'$ are the matrices obtained by term-by-term differentiation of the matrices $B$, $C$ and $F$ respectively with respect to $r$. The corresponding derivations of the formulae obtained for the $\Theta$, $\nu$ and $\rho$ can be found in the Appendix \ref{B}. The sensitivities are usually computed only at $S_0,$ but here we also obtain the sensitivities at underlying prices corresponding to the other grid points, for any $t \in [0,T],$ without significant additional computations.
  
  Efficient computation of sensitivities of derivatives with respect to the underlying risk factors is becoming increasingly important, not only from the perspective of hedging and managing risk, but also to meet certain regulatory requirements. ISDA SIMM \footnote{ INTERNATIONAL SWAPS AND DERIVATIVES ASSOCIATION: ISDA SIMM
TM,1 Methodology, Version 2.0 (December, 2017). \url{https://www.isda.org/a/oFiDE/isda-simm-v2.pdf}} requires the sensitivities of the derivatives for computing the initial margin to be exchanged for over the counter derivatives portfolio. Managing the funding risk associated with the exchange of the initial margin, through the life of the derivative position, involves computing the sensitivities along the simulated scenarios. The proposed method gives a closed form expression of the various sensitivities as a function of time for a grid of underlying values. Following the approach discussed in \cite{kees2}, forward sensitivities of options can be simulated by generating the underlying Monte Carlo scenarios and interpolating the sensitivities obtained at the fixed grid points in the proposed method.

   \section{Numerical Examples}
  \label{nl}
  
  In order to illustrate the performance and efficiency of the method we start with the assumption that the underlying asset price process $S_t$ follows the GBM process and then apply the MOL to compute the value of certain types of European options,  a basic European call option,  a powered option, and finally a cash-or-nothing option; all of which are special cases of the standard Bermudan option defined earlier, where the number of exercise opportunities, $E,$ is set to one. We then report results for a Bermudan put option and numerically study its convergence when an increasing number of spatial grid points are used\footnote{The Python code for all the experiments reported here is available at \url{https://github.com/PurbaBanerjee}}. The following parameters are used throughout for the European styled options: $\sigma=0.3$, $r=0.03$ , initial stock price $S_0 =100$, strike price $K=100$ and $T=1.$ 

Our first step is to generate a uniform mesh $\{\eta_n =n\times dx\}^{N}_{n=0}$ in $[0,1]$ with $dx = 1/N$ and then append an additional point $\eta_{N+1} = 1.1.$ The non-uniform mesh  $\{x_n\}^{N+1}_{n=0}$
is generated using the transformation stated earlier with the value $d = 1.2$ and appropriate values of $c,$ as stated in the respective tables. A brief discussion on choice of the parameter $c$ is provided in Section \ref{paramc}. We  insert in the appropriate position, the point $x= 100,$ in case of the European options, and $x=40,$ in case of the Bermudan put option, and evaluate the accuracy of the method at these points.

 \subsection{European Call option}
 
  As the first numerical test we consider a European call option, whose payoff is given by $\phi(x)=(x-K)^+.$ For the European call option, the closed-form solution of the Black-Scholes equation is given by,
\begin{align*}
    u(\tau,x)=xN(d_1)-K e^{-r\tau}N(d_2),\\
    d_1 =(ln(x/K)+(r+0.5 \sigma^2)\tau)/(\sigma \sqrt{\tau}),~~d_2 =d_1-\sigma\sqrt{\tau}
\end{align*}
where, $N(d)=(1/\sqrt{2\pi})\int_{-\infty}^{d}e^{-0.5 x^2}dx,$ is the cummulative distribution function for the standard normal distribution. The initial conditions for solving this option using the  MOL are:
\begin{align}
\label{ic}
    u(0,x)=\phi(x),
\end{align}
and it also satisfies,
\begin{align}
\label{ic2}
    u(\tau,x)\approx \phi(x)~~\text{as}~~ x\rightarrow \infty
\end{align}

Table \ref{eurocall} reports the values and errors of the European call option obtained by the method when an increasing number of grid points, $N,$ are used. The exact value of the option has been provided in the bracket. 

\begin{table}
\centering
 \begin{tabular}{|c|c|c|}
    \hline
    N & u($13.28330840$) & log error\\
    \hline 100 & 13.25 & -3.7\\
        200 & 13.27 & -5.3 \\
         400 & 13.282 & -6.7\\
         800 & 13.2830 & -8.1\\
         1600 & 13.2832 & -9.7\\
         \hline
    \end{tabular}
     \caption{Value of the European call option obtained by the MOL and the corresponding log errors. We take $c=110,$ the parameter for generating the non-uniform mesh.} \label{eurocall}
\end{table}

\begin{figure}[h]
 \centering
    \includegraphics[width=0.7\textwidth]{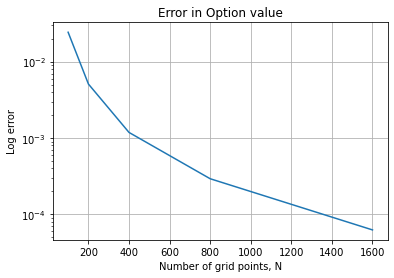}
    \caption{Error in the price of the European call option at the initial point $S_0=100,$ when time to maturity is 1 year and  $c=110$ is used as the parameter for generating the non-uniform mesh.}
    \label{fig:subim1}
    \end{figure}

 Figure \ref{fig:subim1} illustrates the convergence of the method for an increasing number of grid points.  
 
 The Greeks of the European call option are given by \cite{haugcomplete},
\begin{align*}
    \Delta= N(d_1), ~~ \Gamma =\frac{N'(d_1)}{\sigma x \sqrt{\tau}}, ~~ \Theta=-\frac{\sigma x N'(d_1)}{2 \sqrt{\tau}}-r K e^{-r\tau}N(d_2)\\
    \nu =x\sqrt{\tau}N'(d_1),~~ \rho =\tau x e^{-r\tau}N(d_2).
\end{align*}

Table \ref{deltaEU} reports the Delta and Gamma values for the European call option obtained by the MOL along with their respective errors in log scale. We see that fairly accurate values are obtained with relatively few grid points.
\begin{table}
\centering
    \begin{tabular}{|c|c|c|c|c|}
    \hline
    N  & $\Delta(0.59870632)$ & log error($\Delta$) & $\Gamma(0.01288894)$ & log error($\Gamma$) \\
    \hline 100 & 0.6 & -3.3 & 0.0127 & -8.4 \\
        200  & 0.6 & -4.2 & 0.0127 & -9.2 \\
         400  & 0.6 & -5.4 & 0.01285 & -10.4 \\
         800  & 0.6 & -5.4 & 0.01285 & -10.4 \\
         1600 & 0.6 & -6.2 & 0.01287 &  -11.3\\
          \hline
    \end{tabular}
    \caption{Delta and Gamma of the European call option and their corresponding log errors taking $c=110$}  \label{deltaEU}
 \end{table}
 
 Table \ref{vegaEU} reports the vega, rho, and theta values for the European call option. Again we see that fairly accurate values are obtained with relatively few grid points. 
 \begin{table}
 \centering
    \begin{tabular}{|c|c|c|c|c|c|c|}
    \hline
    n   & $\nu (38.667)$ & log error & $\rho(46.5873)$ & log error & $\Theta$ ($-7.19764$) & log error \\
    \hline 100  & 38.7 & -2.4 & 46.6 & -4.0 & -7.2 & -4.3 \\
        200  & 38.68 & -4.1 & 46.59 & -4.9 & -7.2 & -5.9\\
         400  & 38.67 & -5.6 & 46.589& -6.1& -7.198 & -7.4\\
         800   & 38.667 & -7.0 & 46.5878 & -7.5 & -7.1977 & -8.8\\
         1600  & 38.667 & -8.4& 46.5874 & -8.8 & -7.19767 & -10.2\\
         \hline
    \end{tabular}
    \caption{Vega, rho, and theta of the European call option and their corresponding log errors taking $c=110$. }\label{vegaEU}
 \end{table}

\subsection{Powered Option}
 Next , following \cite{jeong2018finite}) we consider a powered option whose pay-off function at maturity $T$ is given by $\phi(x)=$ max$(x-K,0)^p$, where $p$ is a constant (called the power).  The initial conditions of the PDE(\ref{bs2}) for a powered option are given by Equation \ref{icb}, similar to the European call option.

 The closed-form solution of the powered option  is 
$$ u(\tau,x)=\sum_{q=0}^{p}\frac{p!}{q!(p-q)!}x^{p-q}(-K)^q e^{(p-q-1)(r+0.5(p-q)\sigma^2)\tau}N(d_{p,q}),$$
 where, $d_{p,q}=[ln(x/K)+(r+(p-q-0.5)\sigma^2)\tau]/(\sigma\sqrt{\tau})$. In the example  we  choose a value of $p=2$.
 
\begin{figure}[h]
  \centering
    \includegraphics[width=0.7\textwidth]{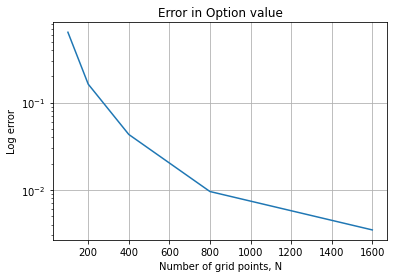}
    \caption{Error in the price of the European powered option at the initial point $S_0=100,$ when time to maturity is 1 year and  $c=123$ is used as the parameter for generating the non-uniform mesh.}
    \label{fig:subim2}
    \end{figure}

Table \ref{power} reports the values of the powered option obtained by the implementation of the MOL for different values of grid points, $N$ and their corresponding log errors. The exact value of the option has been given in the bracket \cite{jeong2018finite}.  Figure \ref{fig:subim2} illustrates the convergence of the powered option with increasing number of grid points. 
 \begin{table}
 \centering
   \begin{tabular}{|c|c|c|}
    \hline
    N & u($676.758$) & log error\\
    \hline 100 & 677.4 & -0.4\\
        200 & 676.9 & -1.8\\
         400 & 676.8 & -3.1\\
         800 & 676.76 & -4.6\\
         1600 & 676.76 & -5.7\\
          \hline
    \end{tabular}
     \caption{Value of the Powered option  obtained by the MOL and it's corresponding log errors taking $c=123$} \label{power}
 \end{table}

Table \ref{deltaGammaP} reports the values and log absolute errors of the values obtained by the MOL, where the reference values (from \cite{jeong2018finite}) are reported in parenthesis.
 \begin{table}
 \centering
  \begin{tabular}{|c|c|c|c|c|}
    \hline
    N  & $\Delta(40.102)$ & log error($\Delta$) & $\Gamma(1.598)$ & log error($\Gamma$) \\
    \hline 100  & 41.2 & 0.1 & 1.592 & -5.1 \\
        200  & 41.1 & 0.1 & 1.592 & -5.1\\
         400  & 41.1 & 0.02 & 1.592 & -5.1 \\
         800  & 40.4 & -1.1 & 1.596 & -6.3\\
         1600  & 40.107 & -5.3 & 1.598 & -7.7\\
         \hline
         \end{tabular}
         \caption{Delta and Gamma of the Powered option and their corresponding log errors taking $c=123$} \label{deltaGammaP}
    \end{table}
 
We also report the vega, rho, and theta values of the powered option in Table \ref{vegaP}. 
 \begin{table}
 \centering
   \begin{tabular}{|c|c|c|c|c|c|c|}
    \hline
    N &  $\nu(4795.291)$& log error($\nu$) & $\rho(3333.420)$& log error($\rho$)& $\Theta$($-819.296$) & log error($\Theta$) \\
    \hline 100  & 4795.4 & -2.4 & 3347.2 & 2.6 & -819.7 & -0.8\\
        200  & 4795.3 & -3.9 & 3336.8 & 1.2 & -819.4 & -2.2 \\
         400  & 4795.293 & -6.3 & 3334.3 & -0.2 & -819.3 & -3.6\\
         800  &  4795.295 & -5.4 & 3333.6 & -1.5 &  -819.3 & -4.9\\
         1600  & 4795.292 & -6.4 & 3333.47 & -2.9  & -819.298 & -6.2\\
         \hline
    \end{tabular}
    \caption{Vega, Rho and Theta of the powered option and their corresponding log errors taking $c=123$} 
    \label{vegaP}
 \end{table}
 
  \subsection{Cash-or-nothing option}
The cash-or-nothing option, with maturity $T$, pays at maturity an amount $C$, provided the value of the underlying asset is greater than $K,$ and no pay-off otherwise. For the experiment, we set $C=K=100$. The exact solution of the cash-or-nothing option is given by:
 \begin{align*}
     u(\tau,x)=Ce^{-r\tau}N(d_2).
 \end{align*}
 
Table \ref{cashTable} reports the values of the cash-or-nothing option obtained for different number of grid points, $N$ and the corresponding log errors. 
  \begin{table}
 \centering
    \begin{tabular}{|c|c|c|}
    \hline
    N & u($46.587$) & log error\\
    \hline 100 & 43.2 & 1.2\\
        200 & 45.5 & 0.1 \\
         400 & 46.0 & -0.6\\
         800 & 46.3 & -1.2\\
         1600 & 46.4 & -1.9\\
          \hline
    \end{tabular}
     \caption{Value of the cash-or-nothing option obtained by the MOL and it's corresponding log errors taking $c=89$}
     \label{cashTable}
 \end{table}
 
 The Greeks of the cash-or-nothing option are given by,
 \begin{align*}
    \Delta=\frac{C e^{-r\tau}N'(d_2)}{\sigma x\sqrt{\tau}},~~\Gamma=-\frac{C d_1  e^{-r\tau}N'(d_2)}{(\sigma x)^2 \tau}\\
     \nu=-C e^{-r\tau}\frac{d_1}{\sigma}N'(d_2),~~\rho=C e^{-r\tau}\left(-\tau N(d_2)+\frac{\sqrt{\tau}}{\sigma}N'(d_2)\right)\\
     \Theta=C e^{-r\tau}\left(rN(d_2)(-\tau N(d_2)+\left(\frac{d_1}{2\tau}-\frac{r}{\sigma \sqrt{\tau}}\right)N'(d_2)\right).
 \end{align*}
 
 The following tables \ref{tab:cash1}, and \ref{tab:cash2} report the values of the Greeks obtained by the MOL along with their respective errors in log scale. The exact values of the respective quantities are given in the brackets (\cite{jeong2018finite}).
 \begin{table}
 \centering
    \begin{tabular}{|c|c|c|c|c|}
    \hline
    N &  $\Delta(1.289)$ & log error($\Delta$) & $\Gamma(-0.011)$ & log error ($\Gamma$)\\
    \hline 100  & 1.27 & -3.9  & -0.007 & -5.5\\
        200  & 1.289 & -5.9 & -0.010 & -6.5\\
         400  & 1.288 & -6.6 & -0.010 & -7.0\\
         800 & 1.288 & -7.0 & -0.010 & -7.4\\
         1600 & 1.288 & -7.4 & -0.011 &  -7.7\\
         \hline
    \end{tabular}
   \caption{Delta and Gamma of the cash-or-nothing option and their corresponding log errors taking $c=89$}
   \label{tab:cash1}
 \end{table}
 
 \begin{table}
 \centering
    \begin{tabular}{|c|c|c|c|c|c|c|}
    \hline
    N &  $\nu(-32.222)$ & log error($\nu$) & $\rho(82.302)$ & log error($\rho$) & $\Theta$($2.364$) & log error($\Theta$) \\
    \hline 100  & -20.7 & 2.4 & 84.4 & 0.7  & 0.5 & 0.6\\
        200   & -28.5 & 1.3 & 83.2 & -0.2 & 1.7 & -0.5 \\
         400   & -30.3 & 0.6 & 82.7 & -0.8  & 2.0 & -1.2\\
         800   & -31.2 & -0.03 & 82.5 & -1.4 &  2.2 & -1.9\\
         1600  & -31.7 & -0.7 & 82.4 & -2.1 & 2.2 & -2.5\\
         
         \hline
    \end{tabular}
    \caption{Vega, Rho and Theta of the cash-or-nothing option and their corresponding log errors taking $c=89$}
    \label{tab:cash2}
 \end{table}
  
\subsection{Bermudan put option} Table \ref{tab:berm} depicts the values, the corresponding log errors and the computational times incurred while implementing the MOL for the Bermudan put option. The reference value for the Bermudan put option is obtained using the COS method \cite{fang2009novel}. The following parameters were used for the asset price model and the option $\sigma=0.3,r=0.06,T=1,$ initial stock price $S_0 =40,$ and the strike $K=44. $ 

 \begin{table}
 \centering
    \begin{tabular}{|c|c|c|c|}
    \hline
    n & u($6.04590214$) & log error & Computational time(seconds)\\
    \hline 250 & 6.0454 & -7.6 & 0.24\\
        500 & 6.0456 & -8.1 & 1.70 \\
         1000 & 6.0458 & -9.8 & 10.83\\
         1500 & 6.0458 & -10.8 & 27.49\\
         2000 & 6.0458 & -11.7 & 67.54\\
         2500 & 6.0458 & -12.6 & 144.86\\
         \hline
    \end{tabular}
    \caption{Value of the Bermudan option obtained by the MOL and it's corresponding log errors along with their computational time. The parameter for non-uniform mesh generation is taken as $c=80.$}
    \label{tab:berm}
 \end{table}
 
 \begin{figure}[h]
  \centering
    \includegraphics[width=0.7\textwidth]{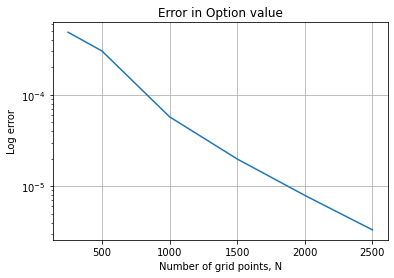}
    \caption{ Error in the price of the Bermudan put option at the initial point $S_0=44,$  a strike of $40,$ when time to maturity is 1 year with $E=10$ equally spaced exercise opportunities. The value of parameter  $c$  for generating the non-uniform mesh is taken as 80.}
    \label{fig:subim3}
    \end{figure}
 
\subsection{Parameter and computational considerations}
\label{paramc}  

\begin{figure}[h]
 \centering
    \includegraphics[width=0.7\textwidth]{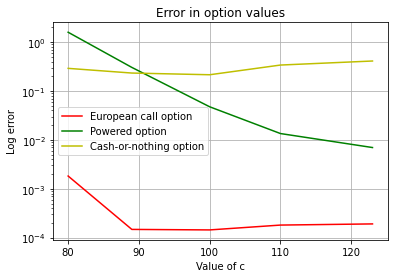}
    \caption{ Error at $S_0=100$ for European call, cash-or-nothing, and powered options when time to maturity is 1 year, with varying values for parameter $c.$}
    \label{fig:subim5}
    \end{figure}

In order to determine the value of parameter $c,$ used for generating the non-uniform mesh, we do a pre-calculation with a relatively small number of grid points $N,$ with varying values of $c.$ Figure \ref{fig:subim5} depicts the changes in the log errors of the European call option, powered option, and cash-or-nothing option for different values of the parameter $c,$ while keeping rest of the parameters constant, and the number of grid points fixed at $N=1000.$

For the European call option we see that the error reduces as we increase the value of $c$ and beyond $c \approx 90,$ the errors are fairly stable. For Powered option increasing $c$ progressively reduces the error, although beyond $c=123,$ there is a slight rise in the error.  For cash-or-nothing the error is stable until $c<100,$ beyond which there seems to be a slight increase in the errors. Based on these heuristics we pre-select the value of $c$ with lowest error for each of the corresponding experiments.

Arriving at the matrices $A$, $B,$ and $C$ in Equation \ref{eq3} for $N$ grid points involves $\mathcal{O}(N),$ computations. The computationally intensive steps involve matrix inversion, which involves $\mathcal{O}(N^3)$ operations, and computing the exponential of the matrix, which again has $\mathcal{O}(N^3)$ operations. Therefore, overall the computational complexity of the method is $\mathcal{O}(N^3).$

\section{Conclusion}
\label{cl}

We have presented an approach based on the MOL for obtaining the value of European and Bermudan options as well as their sensitivities to the various risk factors and model parameters. The MOL approach allows the underlying PDEs to be converted to a system of ODEs, which can be solved using large choice of available efficient solvers. In the Black Scholes framework we show that an exact solution to the ODE can be obtained using exponential of matrix. This makes the presented approach attractive as then we avoid discretization in time, which makes the approach highly efficient. 

For various risk management and regulatory calculations, for instance simulation of future exposure, one has to compute the value of the derivatives along simulated scenarios at various time points. Following the approach discussed in \cite{kees}, efficient solution to the PDE can be combined with Monte Carlo simulations to efficiently compute the exposure along scenarios using interpolation. With the presented approach, one can obtain the exposure of European and Bermudan options for a grid of underlying states, at arbitrary time points. Additionally,  the sensitivities at these spatial grid points can also be computed for any time grid. 

The numerical results obtained for the European options as well as the Bermudan option, along with their respective log errors, serve to prove the desired efficiency of the approach in option valuation. We see that with relatively few spatial grid points, fairly accurate solutions can be obtained.

\bmhead{Acknowledgement}

This paper is dedicated to the memory of Professor Vasudeva Murthy who was instrumental in making this work possible.
 
\medskip
\bibliography{references}

\section{Appendix}
\subsection{Matrix exponential}\label{A}
The matrix exponential is defined for $A\in \mathbb{C}^{n\times n}$ by,
\begin{align}
\label{ea}
    e^A = I + A + \frac{A^2}{2!} + \frac{A^3}{3!}+....
\end{align}
We shall now state an important theorem which yields the convergence of the matrix Taylor series (\ref{ea}).
\begin{thm}(convergence of matrix Taylor series). Suppose $f$ has a Taylor series expansion,
\begin{align}
    f(z) =\sum_{k=0}^{\infty}a_k (z-\alpha)^k, ~~~( a_k =\frac{f^{(k)} (\alpha)}{k!})
\end{align}
with radius of convergence $r$. If $A\in \mathbb{C}^{n \times n}$ then $f(A)$ is defined and is given by
\begin{align}
    f(A)=\sum_{k=0}^{\infty}a_k (A-\alpha I)^k
\end{align}
if and only if each of the distinct eigenvalues $\lambda_1,....,\lambda_s$ of $A$ satisfies one of the conditions
\begin{enumerate}
\item $\lvert \lambda_i -\alpha \rvert < r,$
\item  $\lvert \lambda_i -\alpha \rvert =r,$ 
\end{enumerate}
and the series for $f^{(n_i -1)}(\lambda)$ ( where $n_i$ is the index of $\lambda_i$) is convergent at the point $\lambda =\lambda_i$, $i=1:s$.
\end{thm}
 
 Now, if we consider the power series (\ref{ea}), then the corresponding power series is given by,
 \begin{align}
     f(z)=\sum_{k=0}^{\infty}\frac{z^k}{k!}
 \end{align}
 whose radius of convergence is,
 \begin{align*}
     r=\frac{1}{\lim_{n\rightarrow \infty}\frac{a_{n+1}}{a_n}}
 \end{align*}
 where, $a_n = \frac{1}{n!}$ ( since the limit in the denominator exists in this case). Hence, this readily yields
 \begin{align*}
     r =\frac{1}{\lim_{n\rightarrow \infty} \frac{n}{n+1}}=\infty
 \end{align*}
 Therefore, on applying the theorem above we know that the series (\ref{ea}) is defined and further by standard results in analysis( as in \cite{rudin1964principles}), we can differentiate the series term by term to obtain $\frac{d}{dt}e^{At}=Ae^{At}=e^{At}A.$
 
 Another representation of a matrix exponential is,
 \begin{align}
 \label{ea2}
     e^A =\lim_{s\rightarrow \infty} ( I+A/s)^s
 \end{align}
 The formula above is the limit of the first order Taylor expansion of $A/s$ raised to the power of $s\in \mathbb{Z}$. In a more general setting, we can take the limit as $r\rightarrow \infty$ or $s\rightarrow \infty$ of $r$ terms of the Taylor expansion of $A/s$ raised to the power of $s$, thereby generalising both (\ref{ea}) and (\ref{ea2}). Results show that this general formula too yields $e^A$ and further provides an error bound for finite $r$ and $s$. We shall not delve into these details but refer the reader to \cite{rudin1964principles} for a detailed analysis of the same.
 
\subsection{Derivations of the Greeks}\label{B}
\textbf{Theta($\Theta$)}: Since the matrices do not involve the parameter $t$, a direct differentiation of the function $U(\tau)$ in (\ref{eq4}) with respect to $t$ yields the following formula for $\Theta$,
\begin{align*}
   \frac{\partial U(\tau)}{\partial t} &=-\zeta e^{\tau\zeta}U_0 -{\zeta}^{-1}\zeta e^{\tau\zeta}F\\
    &=-\zeta e^{\tau\zeta}U_0 - e^{\tau\zeta}F
\end{align*}

\textbf{Vega($\nu$):} The calculation of this is a bit involved since the coefficients of the matrices $A$ and $F$ involve the parameter $\sigma$. We first use the identity $\zeta{\zeta}^{-1}=I$ and differentiate it partially with respect to $\sigma$ and utilise the chain rule to obtain,
\begin{align*}
    \frac{\partial \zeta}{\partial \sigma}{\zeta}^{-1}+\zeta\frac{\partial \zeta^{-1}}{\partial \sigma}=0\\
    \implies \zeta\frac{\partial \zeta^{-1}}{\partial \sigma}=-\frac{\partial \zeta}{\partial \sigma}{\zeta}^{-1}\\
     \implies \frac{\partial \zeta^{-1}}{\partial \sigma}=-{\zeta}^{-1}\frac{\partial \zeta}{\partial \sigma}{\zeta}^{-1}
\end{align*}

Now, a simple application of chain rule to (\ref{eq4}) , combined with the formula obtained and the straight-forward result $\frac{\partial \zeta}{\partial \sigma}=A'$ yields the formula (\ref{vega}) for $\nu$.

\textbf{Rho($\rho$):} An almost identical formulation as that of $\nu$ yields the following formula,
\begin{align*}
     \frac{\partial \zeta^{-1}}{\partial r}=-{\zeta}^{-1}\frac{\partial \zeta}{\partial r}{\zeta}^{-1}
\end{align*}
 Using this and the result $\frac{\partial \zeta}{\partial r}=B'+C'$, one readily obtains the formula  (\ref{rho}) for $\rho$.
\end{document}